\documentclass[aps,prl,twocolumn,superscriptaddress,groupedaddress]{revtex4}  

\usepackage{graphicx}  
\usepackage{dcolumn}   
\usepackage{bm}        
\usepackage{amssymb}   
\usepackage[utf8]{inputenc}
\usepackage{color}
\usepackage[multidot]{grffile}
\usepackage{epstopdf}

\begin{document}

\title{SELFAS2 : radio emission from cosmic ray air showers. \\Effect of realistic air refractive index.}

\affiliation{SUBATECH, Universit{\'e} de Nantes \'Ecole des Mines de Nantes IN2P3-CNRS, Nantes France.}

\author{V. Marin$^1$}

\date{\today}
\newcommand{\etal}{\MakeLowercase{\textit{et al. }}} 

\begin{abstract}
Using the simulation code SELFAS2, we present predictions of the radio signal emitted by extensive air showers (EAS) during their development in the atmosphere. The radio emission in the MHz range coming from air showers is the superposition of two mechanisms: the variation of the transverse current due to the systematic opposite drift of electrons and positrons in the Earth's magnetic field and the variation of the charge excess due to the electrons in excess in the shower front. In this paper, we stress particularly the effect of the realistic air refractive index on the radio signal predicted by SELFAS2.
\end{abstract}

\maketitle


\section{introduction}

In 1965, Jelley \textit{et al} \cite{Jelley} and in 1971, Allan \cite{Allan} showed experimentally that the extensive air showers induced by cosmic rays were detectable using the radio detection technique in the MHz range. With the resurgence of this technique in the last decade with the new generation of radio detector arrays like CODALEMA \cite{CODALEMA2005}, LOPES \cite{LOPES}, RAuger \cite{Acounis:2012dg} and recently AERA \cite{AERA}, various modern theoretical approaches to model the radio signal during the air shower development emerged \cite{MGMR,REAS3,SELFAS2,ZHSAIRES}. After more than ten years of modern theoretical studies, the different simulation codes available on the market appear to be in quite well agreement \cite{TIM}, showing that the air shower radio emission is quite well understood. 

The electric field emitted by the air shower during its development is due to two main mechanisms. The first mechanism comes from the time variation of the transverse current caused by the systematic opposite drift of electrons and positrons in the geomagnetic field (the geomagnetic component, see \cite{Allan,CODALEMA2009,LOPES2010} for experimental evidences). The second mechanism comes from the time variation of the net negative charge due to the electrons in excess in the shower front (the charge excess component, see \cite{CORE1,CORE2,Schoorlemmer2012S134} for experimental evidences). 
In most cases, the geomagnetic component is dominant, except for air showers parallel to the magnetic field orientation, where the Lorentz mechanism is weak even null. In this particular case, the proportion of the charge excess component to the total field emitted by the air shower, becomes important even dominant. Recently, it has been shown that the air refractive index plays an important role on the signal amplitude, particularly for antennas located close to the shower axis (less than 300 m) \cite{Cerenkov1, ZHSAIRES}. 

After a brief overview of the code SELFAS2 \cite{SELFAS2}, we will present the latest results predicted by the simulation. We will particularly stress and confirm the influence of a realistic air refractive index on the radio signal emitted by air showers.

\section{SELFAS2 concept}

SELFAS2 is based on a microscopic description of the shower using the concept of "age" and "shower universality" first proposed in \cite{ROSSI} to study the longitudinal development of purely electromagnetic showers. The implementation of relevant distributions for secondary electrons and positrons of the EAS extracted from \cite{LINSLEY,CATALANO,LAFEBRE},
permits to avoid the heavy use of EAS generators to generate air showers in SELFAS2 and makes the simulation completely autonomous. 
The particles generated in SELFAS2 by Monte Carlo simulation, are tracked along their trajectory to compute their individual electric field contribution to the total electric field emitted by the air shower. With this approach, the characteristics of the evolutive spatial density of charge (emissive area) in the shower and the systematic drift of electrons and positrons due to the geomagnetic field are then naturally taken into account in SELFAS2. 

The total electric field as a function of the observer time $t$ seen at the position $\vec{x}$, resulting from the summation of all the particles of the shower, is described in SELFAS2 by:
\begin{eqnarray}
\vec{E}_{tot}(\vec{x},t)=\frac{\eta_i^2}{4\pi  \epsilon _0\epsilon _r}\Bigg\{\sum _{i=1} ^{\zeta}\left[\frac{\vec{n_i}q_i(t_{\mathrm{ret}})}{R_i^2(1-\eta_i\vec{\beta}_i.\vec{n}_i)}\right]_{\mathrm{ret}}\hspace{1.cm}\nonumber\\
\hspace{2.5cm}+\frac{1}{c}\frac{\partial}{\partial t}\sum _{i=1} ^{\zeta}\left[\frac{(\vec{n}_i -\eta_i\vec{\beta}_i)q_i(t_{\mathrm{ret}})}{R_i(1-\eta_i\vec{\beta}_i.\vec{n}_i)}\right]_{\mathrm{ret}}\Bigg\}.
\label{SumField}
\end{eqnarray}
The subscript "ret" means that the quantities inside the bracket $[...]_{\mathrm{ret}}$ have to be evaluated at the retarded time $t_{\mathrm{ret}}$ related to $t$ by the relation of propagation:
\begin{eqnarray}
t=t_{\mathrm{ret}}+\eta_{\mathrm{eff}} \frac{R_i(t_{\mathrm{ret}})}{c}
\label{Time}
\end{eqnarray}
where $R_i$ is the distance between the particle $i$ and the observer. The life time of the particle $i$ is described by $q_i(t_\mathrm{ret})$:
\begin{equation}
q_i(t_{\mathrm{ret}})=\pm e \left[ \Theta(t_{\mathrm{ret}}-t_1^i) - \Theta(t_{\mathrm{ret}}-t_2^i) \right]
\end{equation}
with $\pm e$ the charge of the particle (positron or electron), $t_1^i$, the retarded instant of creation of the moving particle $i$ by sudden acceleration from $\vec{v_i}=0$ to $\vec{v_i}(t_{\mathrm{ret}})$ and $t_2^i$, the retarded instant when the particle $i$ stops by sudden deceleration from $\vec{v_i}(t_{\mathrm{ret}})$ to $\vec{v_i}=0$. The total track length of the particle $i$ contained between $t_1^i$ and $t_2^i$ is subdivided in short tracks along which the velocity is considered as constant. 

\section{Air refractive index and Cerenkov effect}
In the new version of SELFAS2 presented here, the influence of the air refractive index (different from the unity) in the calculation of the signal emitted by the shower, is now taken into account and is described in Eq.\ref{SumField} and in Eq.\ref{Time} by the variables $\eta_i$ and $\eta_{\mathrm{eff}}$ respectively. The value of the air refractive index $\eta_i$ considered at the instant of emission, depends on the altitude of the particle $i$. The model adopted in SELFAS2 to describe $\eta$ as a function of the altitude $h$ (in km)  is described in \cite{INDEX} and is given by: 
\begin{equation}
\eta(h)=1+\eta_{0} \exp(-\alpha h)\times10^{-6}
\label{index}
\end{equation}
where $\eta_{0}=325$ and $\alpha=0.1218$ km$^{-1}$. The variation of the air refractive index with altitude implies that the field propagation from the particle $i$ to the observer at ground is not a straight line, however, it was shown in \cite{INDEXCurved} that the deviation is negligible. As a consequence, the propagation of the field in SELFAS2 is considered as a straight line between the antenna and the observer position. In order to take into account the variation of the air refractive index during the field propagation from the particle $i$ to the observer, the effective air refractive index which has to be considered in \ref{Time} is obtained as follow:
\begin{eqnarray}
n_{\mathrm{eff}}=\frac{1}{h_i-z^{\mathrm{obs}}}\int_{h_i}^{z^{\mathrm{obs}}}n(h)\mathrm{d}h
\label{Neff}
\end{eqnarray}
where $h_i$ is the particle altitude at the instant of emission and $z^{\mathrm{obs}}$, the observer altitude. 
	
The air refractive index different from the unity implies that the denominator in Eq.\ref{SumField} can vanish when the angle $\theta_i$ between the line of sight $\vec{n_i}$ and the direction of the particle propagation $\vec{\beta_i}$, approaches the Cerenkov angle given by:
\begin{equation}
\cos(\theta^{\mathrm{Cer}}_i)=\frac{1}{\eta_i (h_i) \beta_i}.
\end{equation}
Starting again from the integral expression of the electric field given by Eq.(11) in \cite{SELFAS2} and using the fact that in the Fraunhofer approximation, the path difference between the beginning and the end of a short track is given by the distance between these two positions projected onto the line of sight, the electric field emitted by the particle $i$ can be rewritten as: 
\begin{eqnarray}
\vec{E}_i(\vec{x},t)=\frac{\eta_i^2 e^{\pm}}{4\pi  \epsilon _0\epsilon _r}\Bigg\{\frac{\vec{n_i}}{R_i^2}F(t)+\frac{1}{c}\frac{\partial}{\partial t}\left(\frac{(\vec{n}_i -\eta_i\vec{\beta}_i)}{R_i}F(t)\right)\Bigg\}
\end{eqnarray}
with:
\begin{eqnarray}
\hspace{-0.5cm}F(t)=\frac{\Theta[t-\frac{\eta_i R_i}{c}-(1-\eta_i\vec{\beta}_i.\vec{n})t_1]-\Theta[t-\frac{\eta_i R_i}{c}-(1-\eta_i\vec{\beta}_i.\vec{n})t_2]}{1-\eta_i\vec{\beta}_i.\vec{n}_i}
\label{ThetaLim}
\end{eqnarray}
(although we use here the Lorentz gauge, a similar and more detailed demonstration is available in \cite{PhysRevD.81.123009}, using a coulombian gauge). When $\theta_i$ approaches the Cerenkov angle $\theta^{\mathrm{Cer}}_i$, the numerator and the denominator in Eq.\ref{ThetaLim} both vanish. Multiplying and dividing $F(t)$ by $\delta t$, we recognize the first derivative of the Heaviside-step function:
\begin{eqnarray}
F(t)=\delta\left(t-\frac{\eta_i R_i}{c}\right) \delta t
\end{eqnarray}
giving us a limit to estimate the electric field when the observer is located on the Cerenkov angle. 

\section{SELFAS2 results}

\begin{figure}
\includegraphics[scale=0.48]{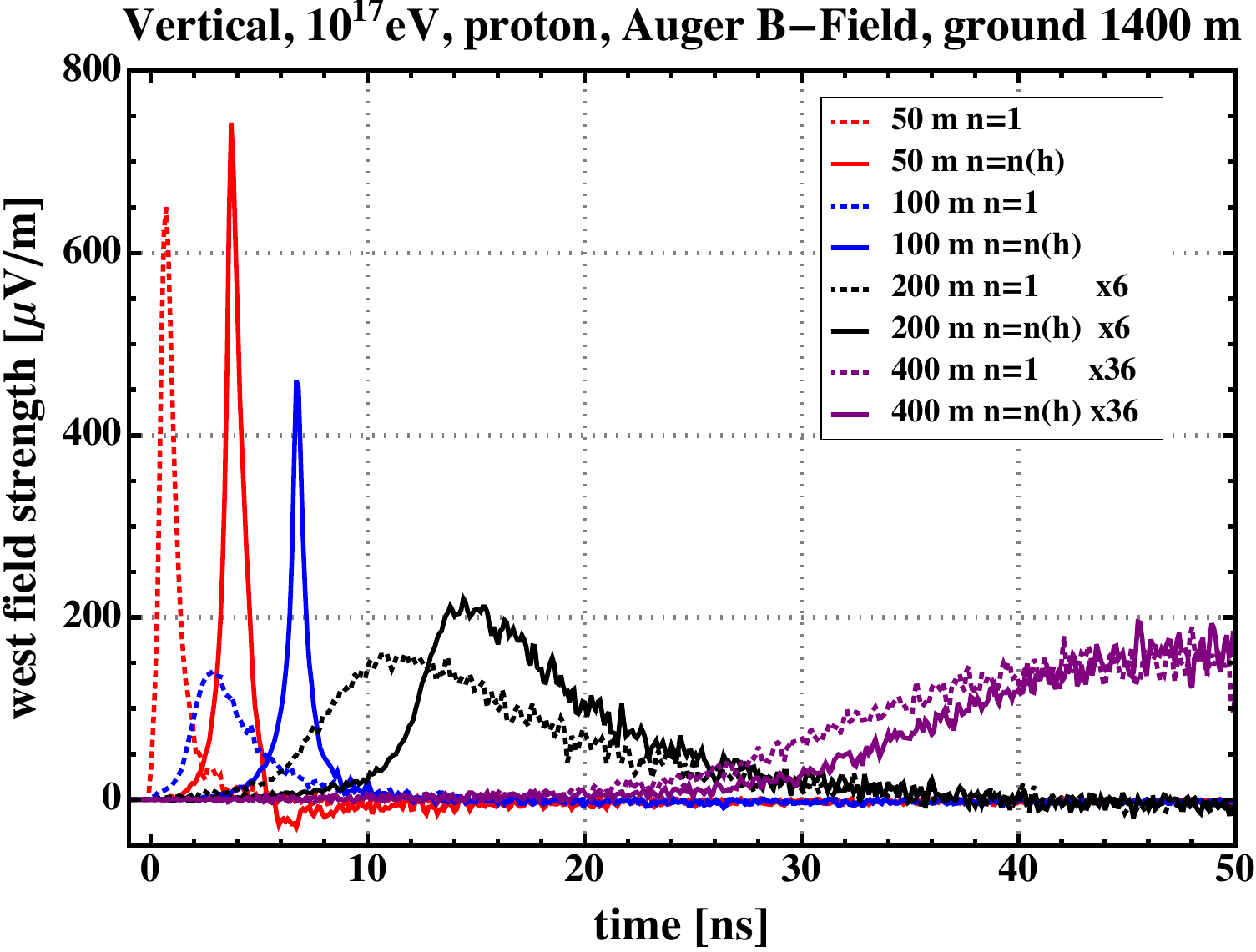}
\caption{\footnotesize{East-west component of the electric field seen by antenna located at different distances from the air shower axis. The dotted lines correspond to the case of an air refractive index fixed to unity ($n=1$), the solid line to the realistic case ($n=n(h)$)}}
\label{Pulses}
\end{figure}

\begin{figure}
\includegraphics[scale=0.37]{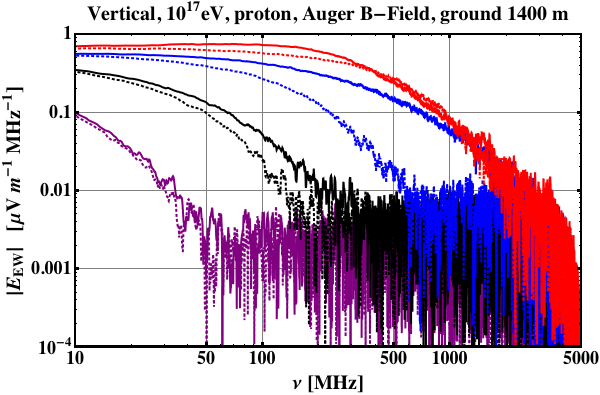}
\caption{\footnotesize{Frequency spectra of pulses shown in Fig.\ref{Pulses} (same legend).}}
\label{Spectras}
\end{figure}

\begin{figure}
\includegraphics[scale=0.51]{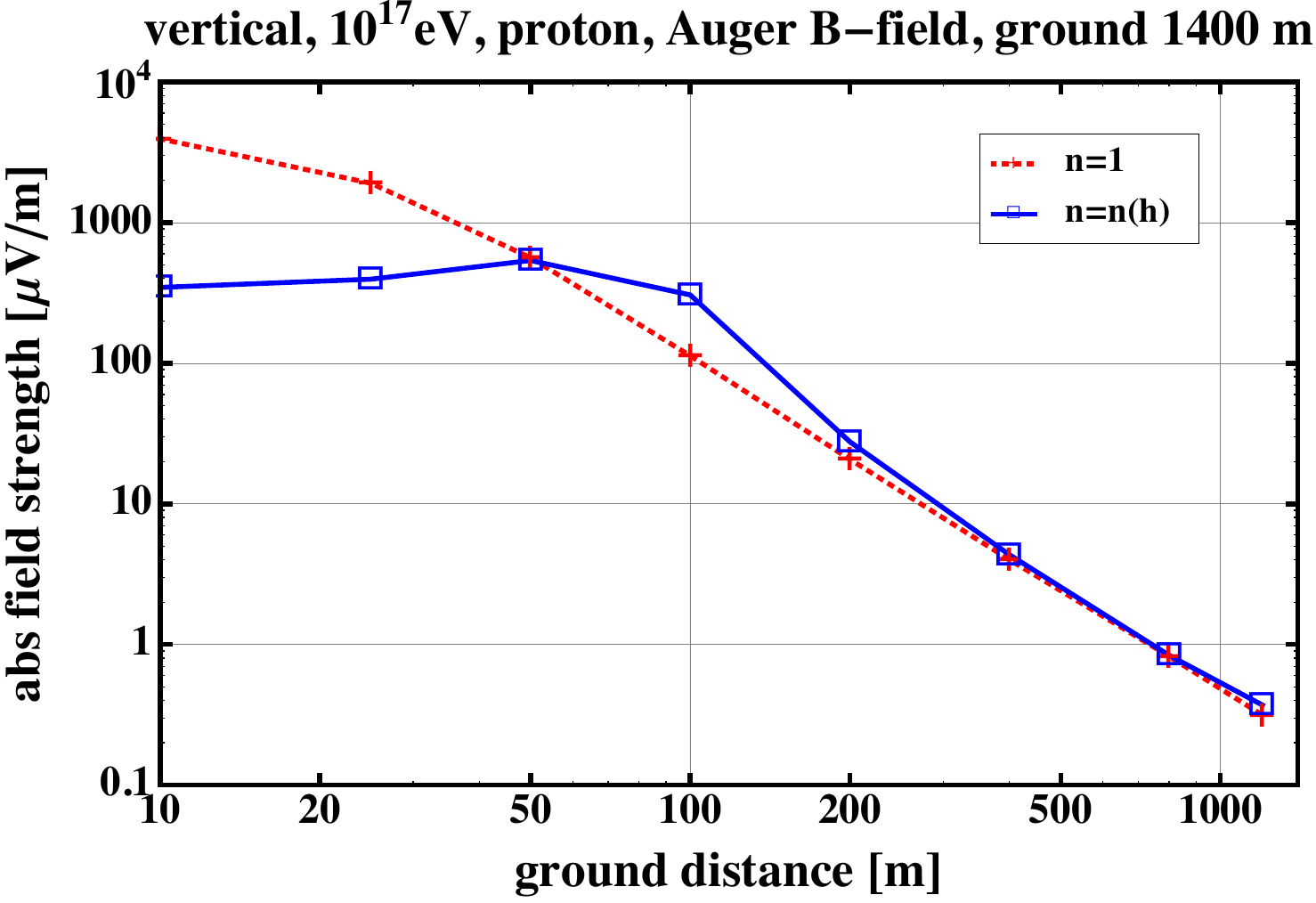}
\caption{\footnotesize{Lateral distribution of the absolute field strength pulses amplitude, observed for antenna located at the north of the ground shower core. The dotted lines correspond to the case of an air refractive index fixe to unity ($n=1$), the full line to the realistic case ($n=n(h)$).}} 
\label{Profils}
\end{figure}

\begin{figure}
\includegraphics[scale=0.3]{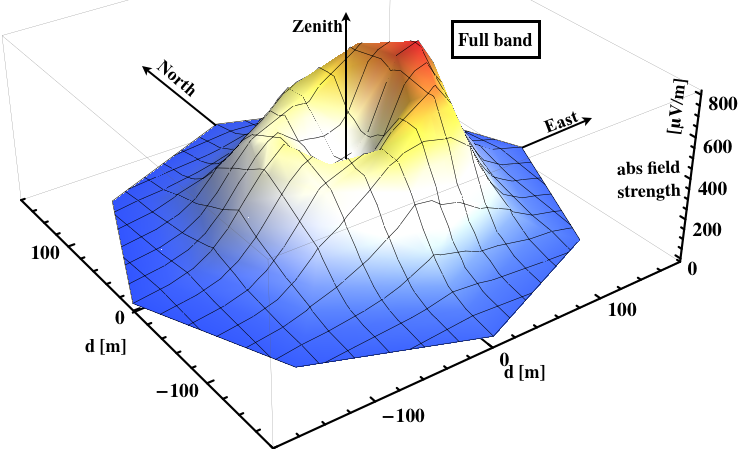}
\caption{\footnotesize{Ground footprint of the east-west polarized radio signal deposited by the air shower (obtained using the maximum of the signal amplitude in the full band).}} 
\label{Ring}
\end{figure}

To underline the effect of the air refractive index, we present in this section the results predicted by SELFAS2 for a vertical air shower induced by a $10^{17}$ eV proton, in the Auger site configuration (geomagnetic field and ground altitude). In Fig.\ref{Pulses} and Fig.\ref{Spectras}, we show the signal observed by antennas located at different distances from the air shower axis, on the east side of the shower core. The results obtained with an air refractive index fixed to unity ($n=1$, dotted lines) are compared to the results obtained with a realistic description of the air refractive index ($n=n(h)$, solid lines). At short axis distance (less, than 300 m), the impact of the air refractive index on the signal amplitude is not negligible, particularly around 100 m. 
The air refractive index influence as a function of the distance to the axis $d$, is not obvious as we see in Fig.\ref{Profils}: the ratio $\frac{E^{n=1}(d)}{E^{n(h)}(d)}$ is not monotonic when $d$ decreases. By analogy with a point-like source moving faster than the speed of light in a medium, the non-monotonic lateral distribution shown in Fig.\ref{Profils} can be considered as an equivalent of the well know "Cerenkov ring" that we can observe in Fig.\ref{Ring}. However for the case of a charge distribution (as it is the case for the shower front) the "ring" observed here is a mix of various effects: 
\begin{itemize}
\item field enhancement when the observation angle approach $\theta_i^{\mathrm{Cer}}$ (see previous section), which is summed up for particles of various energies and directions (due to angular dispersion); 
\item time compression of the signal due to the air refractive index effect on the field propagation (see \cite{Cerenkov1} for instance), which increases the high frequencies amplitude, particularly when the observer is close to the Cerenkov angle, as it is shown in Fig.\ref{Pulses} and Fig.\ref{Spectras}.
\end{itemize}
The air refractive index effect on the radio signal frequently called "Cerenkov effect" is not trivial and should not be compared directly to the classical "Cerenkov radiation". Additionally to the "ring structure" of the radio signal profile at ground, the east-west asymmetry of the radio signal due to the interference between the transverse current contribution and the charge excess contribution (see \cite{MGMR,REAS3,SELFAS2,CORE1} for more details), implies a complex structure of the radio signal observed at ground as we can see in Fig.\ref{Ring}. This shows that caution must be taken with 1-dimensional profile with azimuthal invariance, frequently used for experimental studies.

\begin{figure}
\includegraphics[scale=0.31]{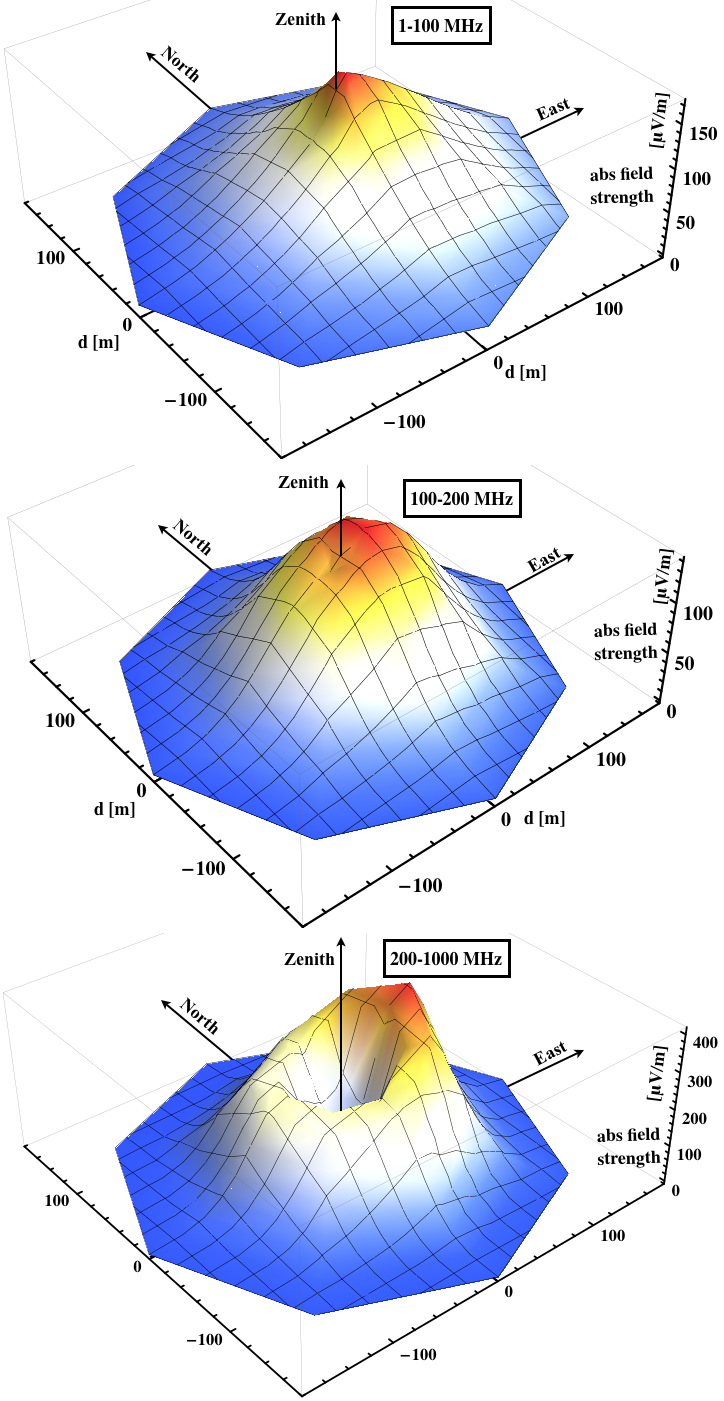}
\caption{\footnotesize{Same as Fig.\ref{Ring} but using the 1-100 MHz band (top), the 100-200 MHz band (center) and the 200-1000 MHz band (bottom).}} 
\label{Footprint1}
\end{figure}

The characterization of the "ring effect" could be of a great interest because its diameter is directly linked to the geometry of the air shower and to the distance between the source and the observer. Assuming that the depth of the maximum of the radio emission is different for a shower initiated by a proton from that initiated by a iron, we can expect that the diameter of the ring could be an observable of the nature of the primary. However, the observation of such effect requires to record the radio signal up to few hundred MHz, to characterize it experimentally as it is shown in Fig.\ref{Footprint1}.

\section{Conclusion}

With this new version of SELFAS2, we confirmed, the importance of a realistic description of the air refractive index to compute the radio signal emitted by air showers. For antennas close to the air shower axis, and more particularly close to the Cerenkov angle, the signal is strongly modified, showing very high frequency component, up to few GHz. Although the effect of the air refractive is stronger for frequencies above 100 MHz, its influence in the experimental frequency windows frequently used (below 100 MHz) by various experiments (as CODALEMA, AERA, RAuger or LOPES for instance), is still important and can enhance the signal by a factor of more than 1.5.


\bibliography{sample}

\end{document}